\documentclass[conference]{IEEEtran}
\IEEEoverridecommandlockouts
\usepackage{cite}
\usepackage{amsmath,amssymb,amsfonts}
\usepackage{graphicx}
\usepackage{textcomp}
\usepackage{xcolor}
\usepackage{listings}
\usepackage{algorithm}
\usepackage{algpseudocode}
\usepackage{tikz}
\usepackage{url}

\usepackage[caption=false]{subfig}
\usepackage{float}
\usepackage{tabularx}
\usepackage{pgf}
\usepackage{xspace}

\def\BibTeX{{\rm B\kern-.05em{\sc i\kern-.025em b}\kern-.08em
    T\kern-.1667em\lower.7ex\hbox{E}\kern-.125emX}}

\newcommand{\name}{{\scshape VeilGraph}\xspace}
\newcommand{\titlename}{VeilGraph\xspace}

\definecolor{dkgreen}{rgb}{0,0.6,0}
\definecolor{gray}{rgb}{0.5,0.5,0.5}
\definecolor{mauve}{rgb}{0.58,0,0.82}

\def\BibTeX{{\rm B\kern-.05em{\sc i\kern-.025em b}\kern-.08em
    T\kern-.1667em\lower.7ex\hbox{E}\kern-.125emX}}
    
\newcommand*\annotatedFigureText[4]{\node[draw=none, anchor=south west, text=#2, inner sep=0, text width=#3\linewidth,font=\sffamily] at (#1){#4};}
\newenvironment {annotatedFigure}[1]{\centering\begin{tikzpicture}
\node[anchor=south west,inner sep=0] (image) at (0,0) { #1};\begin{scope}[x={(image.south east)},y={(image.north west)}]}{\end{scope}\end{tikzpicture}}

\begin{document}

\title{\vspace{-3pt}VeilGraph: Approximating Graph Streams\\
{\footnotesize \textsuperscript{*}This document merely serves the purpose of timely dissemination. Copyrights belong to original holders.}
\vspace{-7pt}}


\author{\IEEEauthorblockN{Miguel E. Coimbra}
\IEEEauthorblockA{\textit{INESC-ID/IST} \\
\textit{Universidade de Lisboa}\\
Lisbon, Portugal \\
miguel.e.coimbra@tecnico.ulisboa.pt}
\and
\IEEEauthorblockN{Sérgio Esteves}
\IEEEauthorblockA{\textit{Feedzai/INESC-ID/IST} \\
\textit{Universidade de Lisboa}\\
Lisbon, Portugal \\
sesteves@gsd.inesc-id.pt}
\and
\IEEEauthorblockN{Alexandre P. Francisco}
\IEEEauthorblockA{\textit{INESC-ID/IST} \\
\textit{Universidade de Lisboa}\\
Lisbon, Portugal \\
aplf@inesc-id.pt}
\and
\IEEEauthorblockN{Luís Veiga}
\IEEEauthorblockA{\textit{INESC-ID/IST} \\
\textit{Universidade de Lisboa}\\
Lisbon, Portugal \\
luis.veiga@inesc-id.pt}
\vspace{-7pt}}


\maketitle

\begin{abstract}
Graphs are found in a plethora of domains, including online social networks, the World Wide Web and the study of epidemics, to name a few.
With the advent of greater volumes of information and the need for continuously updated results under temporal constraints, it is necessary to explore novel approaches that further enable performance improvements.

In the scope of stream processing over graphs, we research  the trade-offs between result accuracy and the speedup of approximate computation techniques.
We see this as a natural path towards these performance improvements.
Herein we present \name, through which we conducted our research.
We showcase an innovative model for approximate graph processing, implemented in \texttt{Apache Flink}.

We analyze our model and evaluate it with the case study of the PageRank algorithm~\cite{pageRank}, perhaps the most famous measure of vertex centrality used to rank websites in search engine results.
Our experiments, even when set up for favoring \texttt{Flink} for comparability,  show that \name can improve performance up to 3X speedups, while achieving result quality above 95\% when compared to results of the traditional version of PageRank without any summarization or approximation techniques.
\end{abstract}


\section{Introduction}\label{sec:intro}

Working with large graphs that are continuously changing in real-time, with a stream of unbounded updates, is an increasingly important and challenging problem.
Not only the graph topology is constantly changing with the addition and removal of edges and vertices, but also query response times must be interactive and meet stringent latency demands.
Common domains of applicability of such graphs include social networks, recommendation systems, and people and vehicle position tracking.
In these domains, the ability to quickly react to change would allow for useful detection of trends.

Graph Processing Engines (GPEs) often resort to approximate computing techniques in order to provide timely query responses in very large graphs, without adding extra resources.
Approximate computing allows inaccurate query results in exchange of lower latencies.
Under specific error bounds, approximate results would be as equally acceptable as the exact answers for many scenarios.
Approximate results may allow for considerable improvements in speed (e.g., reduced latency and processing time, increased throughput) and resource efficiency (e.g., reducing cloud computing costs and energy footprint).
In this domain, three common techniques have been employed: \emph{sampling}, where queries are executed on a sampled summarization of the graph~\cite{GraphSampling2013}; \emph{task dropping}, which consists of discarding parts of a partitioned global task processing list~\cite{ProbabilisticAccuracy2006}; and \emph{load shedding}, which partially discards inputted data according to a shedding scheme~\cite{LoadShedding2003}.
%
%
Developing novel techniques for approximate graph processing can strongly benefit many systems and applications.
This would pave the road for high-level optimizations like Service-Level Agreements (SLAs) for graph processing, with different tiers of accuracy and resource efficiency.
These are relevant for applications like product recommendation and monitoring user influence.

We introduce \name, a novel execution model for GPEs that enables approximate computations on general graph applications.
Our model features an abstraction that flexibly allows the expression of custom vertex impact estimators for random walk based algorithms.
With this abstraction, we build a representative graph summarization that solely comprises the subset of vertices estimated as yielding high impact.
This way, \name is capable of delivering lower latencies in a resource-efficient manner, while maintaining query result accuracy within acceptable limits. 
As a concrete instance, we integrated \name with a modern and popular GPE, \texttt{Gelly/Apache Flink}.
Experimental results indicate that our approximate computing model can achieve a 2-fold latency improvement over the base (exact) computing model, while not degrading result accuracy in more than 5\%.


The rest of the paper is organized as follows.
An overview of the \name model is provided in Section~\ref{sec:graphbolt-model}.
Section~\ref{sec:graphbolt-arch} describes the architecture.
In Section~\ref{sec:evaluation} we present the experimental evaluation, followed by an analysis of improvements.
Section~\ref{sec:related} analyzes state-of-the-art GPEs that approach similar challenges.
We summarize our contribution and future research in Section~\ref{sec:conclusion}.


\section{Model: Big Vertex}\label{sec:graphbolt-model}

We explicitly separate the graph algorithm expression paradigm and the underlying summarization model.
Allowing for different graph summarization models, the goal is to enable different approximate computation strategies in exchange for result accuracy.
In this work, we implemented and analyzed what we call the \textit{big vertex} model.
When processing a stream of new edges, the parameters of our model highlight a subset $K$ of the graph's vertices, known as \textit{hot vertices}.
The aim of this set is to reduce the number of processed vertices as close as possible to $O(K)$.
These vertices are used to update the algorithm output.

\subsection{Motivation: Not all vertices are equal}\label{subsec:not_all_vertices}
In order to use only a subset $K$ of the vertices, it is necessary to employ approximate computing techniques using a fraction of the total data.
In this model there is an aggregating vertex $\mathcal{B}$.
We refer to $\mathcal{B}$ as the \textit{big vertex} -- a single vertex representing all the vertices outside $K$ (in this model, the values are not updated for vertices in $\mathcal{B}$).
For the original graph $G = (V, E)$ and vertex set $K$, we define a summary graph $\mathcal{G} = (\mathcal{V}, \mathcal{E})$, where $\mathcal{V} = K \cup \{\mathcal{B}\}$.
We define $\mathcal{E} = E_{K} \cup E_{\mathcal{B}}$, where $E_{K} = \{ (u, v) \in E: u, v \in K\}$, which is the set of edges with both source and target vertices contained in $K$ and $E_{\mathcal{B}} = \{ (w, z) \in E : w \not \in K, z \in K \}$ as the set of edges with sources contained \textit{inside} $\mathcal{B}$ and target in $K$.
Conceptually, this consists in replacing all vertices of $G$ which are not in $K$ by a single big vertex $\mathcal{B}$ and representing the edges whose targets are \textit{hot vertices} and whose sources are now in $\mathcal{B}$.
The summary graph $\mathcal{G}$ does not contain vertices outside of $K$ (again, those are represented by $\mathcal{B}$).

It is relevant to retain that for each iteration, the impact of a vertex $v$ depends on what is received through its incoming edges.
By definition, $\mathcal{B}$ represents all vertices whose impact is not expected to change significantly.
The contribution of each vertex $v \not \in K$ (and therefore represented by $\mathcal{B}$) is constant between iterations, so it can be registered initially and used afterward.
As a consequence, the summary graph $\mathcal{G}$ does not contain edges targeting vertices represented by $\mathcal{B}$.
However, their existence must be recorded: even if the edges coming out of $K$ and into $\mathcal{B}$ are irrelevant for the computation, they still matter for the vertex degree, which influences the emitted scores of the vertices in $K$.
Despite the fact those edges targeting $\mathcal{B}$ are being discarded when building $\mathcal{G}$, the summarized computation must occur as if nothing had happened.
To ensure correctness, for each edge $(u, v) \in E_{K}$, we store $val((u, v)) = 1 / d_{out}(u)$ with $d_{out}(u)$ as the out-degree of $u$ before discarding the outgoing edges of $u$ targeting vertices in $\mathcal{B}$.

It is also necessary to record the contribution of all the vertices fused together in $\mathcal{B}$.
For each edge whose source $w$ is inside $\mathcal{B}$ and whose target $z$ is in $K$, we store the contribution that would originally be sent from $w$ as $val((w, z)) = w_{s} / d_{out}(w)$ where $w_{s}$ is the stored score/value of $w$ and the out-degree of $w$ is defined as $d_{out}(w)$.
The contribution of $\mathcal{B}$ as a single vertex in $\mathcal{G}$ is then represented as $\mathcal{B}_{s}$ and defined as:
\begin{equation}
\mathcal{B}_{s} = \sum_{w\in\mathcal{B}} \sum_{(w,z)\in\ E_\mathcal{B}} val((w,z))
\end{equation}
The fusion of vertices into $\mathcal{B}$ is performed while preserving the global influence from vertices placed inside $\mathcal{B}$ to vertices in $K$.
Our model intuition is that vertices receiving more updates have a greater probability of having their measured impact change in between execution points. 
Their neighboring vertices are also likely to incur changes, but as we consider vertices further and further away from $K$, contributions are likely to remain the same~\cite{LinkEvolutions, LoadShedding2003}. 

\name aims to enable approximate computing on a stream $S$ of incoming edge updates. 
While we focused on the class of random walk algorithms, \name's architecture has the potential to be applied to other classes of graph algorithms (and potentially, other models, depending on the relation between a given graph algorithm and the summarization model) using the same principles.
Our contribution strikes a balance between two strategies for when a query is to be served: \textit{a)} recomputing the whole graph properties when a query arrives; \textit{b)} returning a previous query result without incurring any type of additional computation.
While the former is obviously much more time-consuming, it has the property of maintaining result accuracy.
The latter, on the other hand, would quickly lower the accuracy of the algorithm's results.
Throughout this document, we use the term \textit{query} to state that a graph algorithm's results are required.
So when it is said that a query is executed, it means that the algorithm was executed, independently of being executed over the complete graph or our summarization model.

\subsection{Design: Building the model}\label{subsec:building_the_model}

The model is based on techniques such as defining and determining a confidence threshold for error in the calculation~\cite{Knowing2014}, graph sampling~\cite{GraphSampling2013} and other hybrid approaches.
\name registers updates as they arrive for both statistical and processing purposes.
Vertex and edge changes are kept until updates are formally applied to the graph.
Until they are applied, statistics such as the total change in the number of vertices and edges (with respect to accumulated updates) are readily available.
When applying the generic concepts of our technique, a useful insight is that most likely, not all vertex scores will need to be recalculated.
In our \textit{big vertex} model, generating the subset $K$ of a graph $G =\ $($V, E$) depends on three parameters ($r, n, \Delta$) used in a two-step procedure.
From the client perspective, we consider a query to be an updated view of the algorithm information pertaining $G$.
As each individual query represents an important instant as far as computation is concerned, we refer to each query as a measurement point $t$.
For any measurement point $t$, the whole graph is represented as $G_{t} =\ $($V_{t}, E_{t}$).

\begin{figure}
    \centering

    \begin{annotatedFigure}
        {\includegraphics[width=0.990\linewidth]{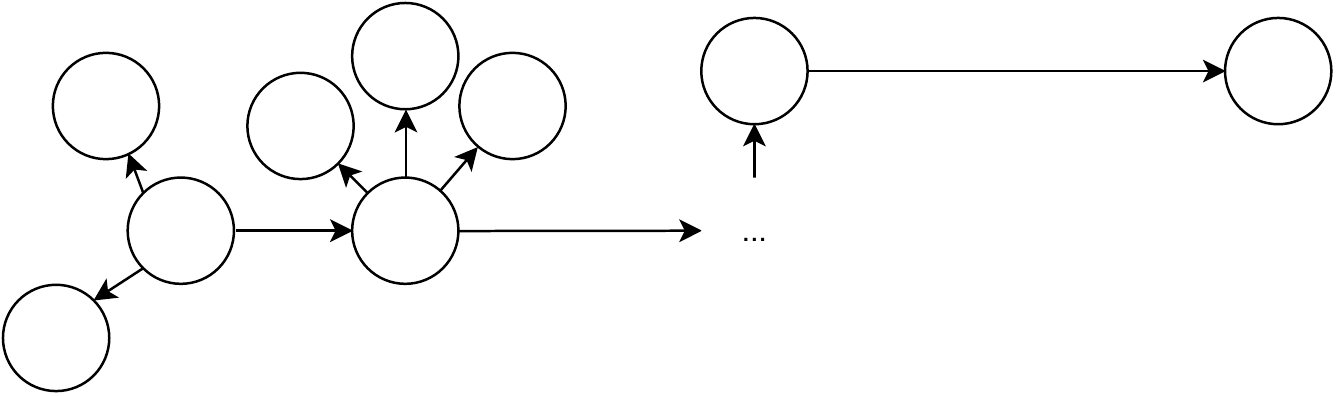}}
        \annotatedFigureText{0.122,0.36155}{black}{0.5}{$v$}
        \annotatedFigureText{0.162,0.06155}{black}{0.5}{$\dfrac{v_{s}}{d_{t}(v)}$}
        \annotatedFigureText{0.292,0.36155}{black}{0.5}{$u$}
        \annotatedFigureText{0.33,0.06155}{black}{0.5}{$\dfrac{v_{s}}{d_{t}(v)d_{t}(u)}$}
        \annotatedFigureText{0.552,0.7435}{black}{0.5}{$j$}
        \annotatedFigureText{0.607,0.47155}{black}{0.5}{$\dfrac{v_{s}}{d_{t}(v)d_{t}(u)\dots\ d_{t}(j)}$}
        \annotatedFigureText{0.947,0.7535}{black}{0.5}{$i$}
    \end{annotatedFigure}

    \caption{The contribution of vertex $v$ diminishes as we expand further away from it (Eq.~\ref{eq:delta_function}). Parameter $\Delta$ dictates how far to expand around vertices until the accumulated fraction drops below $\Delta$. $v_{s}$ is the score of vertex $v$; $d_{t}(v)$ is the out-degree of vertex $v$ at measurement point $t$.}

    \label{fig:delta_intuition}
\end{figure}
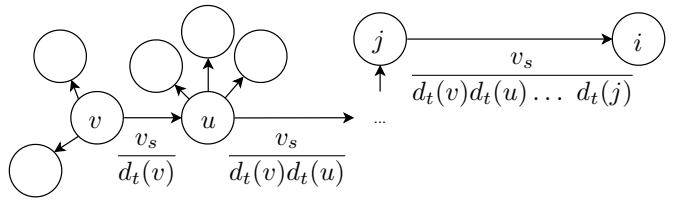

\begin{enumerate}

\item \textbf{Update ratio threshold \textit{r}.}
This parameter defines the minimal amount of change in a vertex $u$'s degree in order to include $u$ in $K$.
Parameters $r$ and $n$ are parameters of \name's \textit{big vertex} model to harness a graph algorithm's heuristics to approximate a result.
We adopt the notation where the set of neighbors of vertex $u$ in a directed graph at measurement instance $t$ is written as $N_{t}\left(u\right)=\{v \in V: \left(u, v\right)\in E_{t}\}$.
We further write the degree of vertex $u$ in measurement instance $t$ as $d_{t}\left(u\right) = |N_{t}\left(u\right)|$.
The function $d\left(u, v\right)$ represents the length (number of hops) of the minimum path between vertices $u$ and $v$ and $d_{t}\left(u, v\right)$ represents the same concept at measurement instance $t$.
It is not required to maintain shortest paths between vertices (that would be a whole different problem~\cite{kalavri2016shortest}).
This model is based on a vertex-centric breadth-first neighborhood expansion.
Let us define as $K_{r}$ the set of vertices which satisfy parameter $r$, where $d_{t}\left(u\right)$ is the degree of vertex $u$, $t$ represents the current measurement point and $t - 1$ is the previous measurement point:\footnote{New vertices are always included in $K$. The subtraction in the formula registers the degree change ratio with respect to the previous value $d_{t - 1(u)}$.}

\begin{equation}\label{eq:r}
K_{r}=\left\{u: \left|\frac{d_{t}\left(u\right)}{d_{t - 1}\left(u\right)}\ - 1\right| > r\right\}
\end{equation}

\item \textbf{Neighborhood diameter \textit{n}.}
This parameter is used as an expansion around the neighborhood of the vertices in $K_{r}$.
It aims to capture the locality in graph updates: those vertices neighboring the ones beyond the threshold, and as such still likely to suffer relevant modifications when vertices in $K$ are recalculated (attenuating as distance increases).
On measurement point $t$, for each vertex $u \in K_{r}$, we will expand a neighborhood of size $n$, starting from $u$ and including every additional vertex $v \in V_{t} \setminus K_{r}$ found in the neighborhood diameter expansion.
The expansion is then defined as:

\begin{equation}\label{eq:n}
K_{n}=\{v: d_{t}\left(u, v\right) \leq n, u \in K_{r}, v \in V_{t} \setminus K_{r}\}
\end{equation}
$V_{t}$ is the set of vertices of the graph at measurement point $t$.
$n = 0$ is set to promote performance, while a greater value of $n$ is expected to trade performance for more accuracy.

\item \textbf{Result-specific neighborhood extension $\Delta$.}
\label{misc:summarygraph}
This last parameter allows users to extend the functionality of $n$ by further expanding neighborhood size as a function of vertices' results.
This is achieved by accounting for specific underlying algorithm's properties which may be used as heuristics.
This allows updating vertex results around those vertices that, while not included by Eq.~\ref{eq:r} or Eq.~\ref{eq:n}, are neighbors to vertices subject to change.
We use the relative change of vertex score between the two consecutive measurement points $t-1$ and $t$.
The expansion of the neighborhood with $\Delta$ is:
\begin{equation}
\begin{split}
K_{\Delta} = \{v: d_{t}\left(u, v\right) \leq f_{\Delta}\left(v\right),\\ u \in K_{n}, v \in V_{t} \setminus \{K_{r}\cup K_{n}\}\}
\end{split}
\end{equation}
where $f_{\Delta}\left(v\right)$ is the $\Delta$-expansion function:

\begin{equation}\label{eq:delta_function}
f_{\Delta}\left(v\right) = \frac{1}{\log\overline{d}}\,  \log \left(\dfrac{\overline{d} \, v_{s}}{\Delta \, d_{t}\left(v\right)} \right)
\end{equation}

In Eq.~\ref{eq:delta_function}, $v_{s}$ is a result 
on vertex $v$ and $\overline{d}$ is the average degree of the currently accumulated vertices with respect to stream $S$.
This allows us to have a fine-grained neighborhood expansion starting at $v$ and limited by $\Delta$ on the maximum contribution of the score of $v$.
The intuition here is that vertex $v$ would contribute to the value of its immediate neighbors with a value of $\frac{v_{s}}{d_{out}(v)}$.
For its neighbors' neighbors,  the influence of $v$ would be further diminished (contribution would now be $\frac{v_{s}}{d_{out}(v)d_{out}(u)}$ where $u$ is a direct neighbor of $v$. 
Additional expansions would further dilute the contribution that $v$ could possibly have.
For example, when evaluating \name with a bound of $\Delta = 0.1$, we keep considering further neighborhood expansion hops from $v$ until the contribution from $v$ drops below 10\% of its score.
Figure~\ref{fig:delta_intuition} provides a visual example of the dilution of the score of vertex $v$ due to the degree of the vertex at the end of each hop ($u$, then intermediate vertices represented as '\dots', followed by $j$ until $i$ is reached).
\end{enumerate}

\begin{figure}[h!t]
\centering
\begin{annotatedFigure}
{\includegraphics[width=0.95\linewidth]{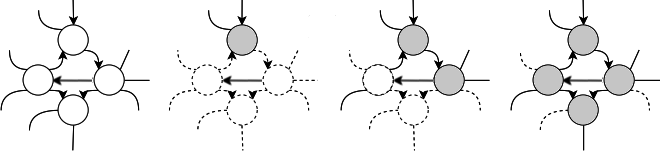}}
	\annotatedFigureText{0.101,-0.2435}{black}{0.1}{$\emptyset$}
	\annotatedFigureText{0.251,-0.2435}{black}{0.1}{a)\\$K_{r}$}
	\annotatedFigureText{0.504,-0.2435}{black}{0.3}{b)\\$K_{r} \cup K_{n}$}
	\annotatedFigureText{0.748,-0.2435}{black}{0.5}{c)\\$K_{r} \cup K_{n} \cup K_{\Delta}$}
\end{annotatedFigure}
\caption{Evolution of $K$ in the \name model.}
\label{fig:params}
\end{figure}
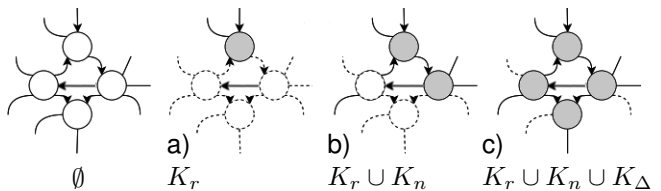

We then have a set of \textit{hot vertices} $K = K_{r} \cup K_{n} \cup K_{\Delta}$ which is used as part of a graph summary model (deriving from techniques in iterative aggregation~\cite{Langville:2004:UPI:1013367.1013491}), written as $\mathcal{G} = (\mathcal{V}, \mathcal{E})$.
An example of how the parameters influence the selection of vertices is depicted in Figure~\ref{fig:params}.
The left side represents a zoom of a small portion of the complete graph $G$ (which may be composed of millions of vertices).
Part \textit{a)} shows the vertices whose amount of change satisfied the threshold ratio $r$, leading to $K_{r}$: in this case, only the top vertex was included, for which its contour is a solid line with a gray fill.
Part \textit{b)} then shows the usage of the neighborhood expansion parameter $n=1$ over the vertex of \textit{a)}: the two middle vertices are now gray as well (the one included in the previous part is colored black).
This makes up $K_{r} \cup K_{n}$.
Lastly, part \textit{c)} accounts for the per-vertex neighborhood expansion parameter $\Delta$ and represents the set of \textit{hot vertices} $K = K_{r} \cup K_{n} \cup K_{\Delta}$.
This is depicted by coloring the bottom vertex in gray and the remaining ones in black (they are already part of $K$).
Some dashed arrows remain to illustrate that the vertices on the other end of the edges were not included in $K$.

\begin{algorithm*}[ht]
\floatname{algorithm}{Algorithm}
\fontsize{8}{9}\ttfamily
\caption{\name Execution Skeleton:\quad graph G, stream S}\label{alg:approx-proc}
\begin{algorithmic}[1]
\State \Call{OnStart}{G, S} \quad /* Initializations. */
\State graphUpdates $\gets \varnothing$ \quad \State updateStatistics $\gets \varnothing$
\Repeat
    \State msg $\gets$ \Call{TakeMessage}{S}
    \If{msg is \textit{Add}}
    	\Call{RegisterAddEdge}{msg, graphUpdates, updateStatistics}
    \ElsIf{msg is \textit{Remove}}
    	\Call{RegisterRemoveEdge}{msg, graphUpdates, updateStatistics}
    \ElsIf{msg is \textit{Query}}
    	\State needToApplyUpdates? $\gets$ \Call{CheckUpdateState}{graphUpdates, updateStatistics}
        \If{needToApplyUpdates?}
        	\State G $\gets$ \Call{ApplyUpdates}{graphUpdates, updateStatistics}
        \EndIf
        \State strategy $\gets$ \Call{DecideQueryStrategy}{msg, G, updates, updateStatistics}
        \If{strategy = \textit{Repeat-last-answer}}
            \State newResults $\gets$ previousResults
        \ElsIf{strategy = \textit{Compute-approximate}}
            \State newResults $\gets$ \Call{ComputeApproximate}{G, previousResults}
        \ElsIf{strategy = \textit{Compute-exact}}
            \State newResults $\gets$ \Call{ComputeExact}{G}
        \EndIf
        \State \Call{OutputResults}{newResults}
        \State \Call{OnQueryResult}{msg, G, newResults, jobStatistics}\quad /* Extrapolate and store job statistics. */
    \EndIf
\Until{stopped}
\State \Call{OnStop}{ }\quad /* Tear-down procedure. */

\end{algorithmic}
\end{algorithm*}


\section{\name Architecture}\label{sec:graphbolt-arch}

\begin{figure}
\centering
\includegraphics[width=\linewidth]{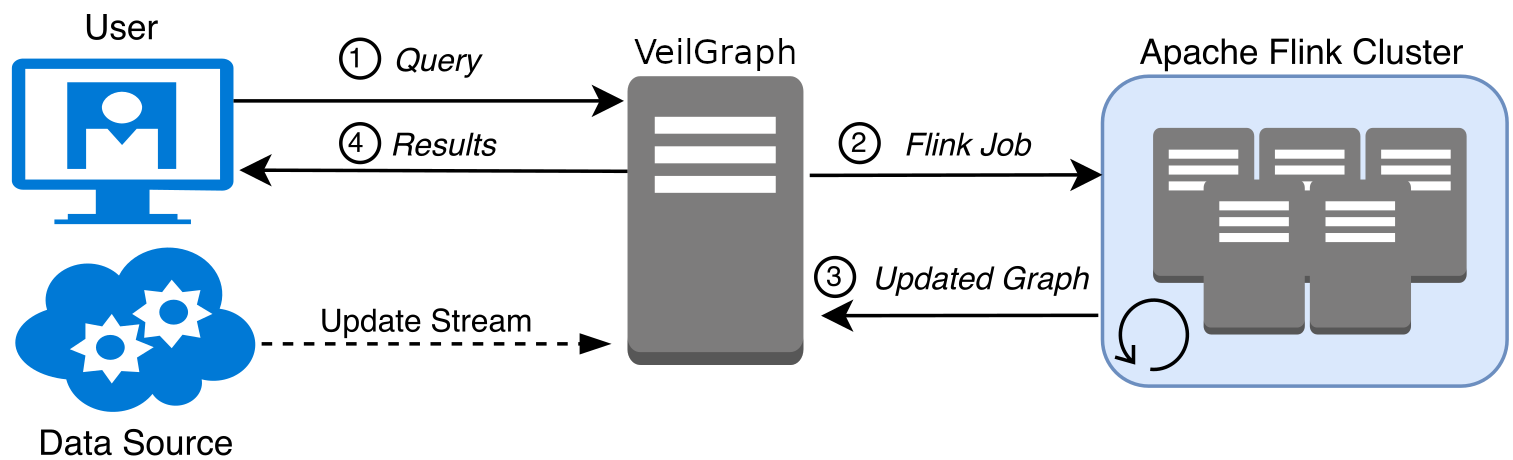}
\caption{Diagram of \name workflow on a \texttt{Flink} cluster.}
\label{fig:graphbolt-usecase}
\end{figure}

The architecture of \name was designed while taking in account three information types which are inherent to the work flow, presented in Figure~\ref{fig:graphbolt-usecase}.
The \name module will constantly monitor one or more streams of data and track the changes made to the graph.
When the data is \textit{queried}, \name will execute the request by submitting a job to a \texttt{Flink} cluster.
In our experiments, we trigger the incorporation of updates into the graph whenever a client query arrives.\footnote{While outside the scope of this work, a live scenario would have a more elaborate ingestion scheme, possibly using dedicated ingestion nodes like in KineoGraph~\cite{Cheng:2012:KTP:2168836.2168846}.}
The main elements of the flow of information in an execution are:

\begin{itemize}

\item \textbf{Initial graph \textit{G}}.
The original graph upon which updates and queries will be performed.

\item \textbf{Stream of updates \textit{S}}.
Our model of updates could be the removal $e_{-}$ or addition $e_{+}$ of edges and the same for vertices ($v_{-}, v_{+}$).
We make as little assumptions as possible regarding $S$: the data supplied needs not respect any defined order.
In our experiments we used both edge additions and removals.

\item \textbf{Result \textit{R}}.
Information produced by the system as an answer to the queries received in $S$.
\end{itemize}

\name was designed to allow programmers to define fine-grained logic for the approximate computation when necessary.
This is achieved through the usage of user-defined functions.
As a design decision, there are five distinct functions which are articulated in a fixed program structure.
The API of \name uses them to define the execution logic that will guide the processing strategy.
They are key points in execution where important decisions should take place (e.g., how to apply updates, how to perform monitoring tasks).
To implement other graph algorithms, users can simply extend the \name \texttt{Java} class implementing the logic shown in Algorithm~\ref{alg:approx-proc} to enable our model's functionality, while abstracting away many implementation details (inherent to our architecture) unrelated to the graph processing itself.

Additional behavior control is possible by customizing the model by implementing their own functions (left as \texttt{abstract} methods of the class implementing the architecture logic).
Overall, this approach has the advantage of abstracting away the API's complexity, while still empowering power users who wish to create fine-tuned policies.
\name's architecture creates a separation between the graph model, the way the graph processing is expressed (e.g. such as vertex-centric) and the function logic to apply on vertices.

\name was implemented over \texttt{Apache Flink}~\cite{Junghanns:2016:AEP:2980523.2980527}, a framework built for distributed stream processing.\footnote{\url{https://flink.apache.org/}}
It has many different libraries, among which \texttt{Gelly}, its official graph processing library.
It features algorithms such as PageRank, Single-Source Shortest Paths and Community Detection, among others.
Overall, it empowers researchers and engineers to express graph algorithms in familiar ways such as the \textit{gather-sum-apply} or the \textit{vertex-centric} approach of Google Pregel~\cite{Malewicz:2010:PSL:1807167.1807184}, while providing a powerful abstraction with respect to the underlying scheme of distributed computation.
We employ \texttt{Flink's} mechanism for efficient dataflow iterations~\cite{Kalavri:2014:ALG:2621934.2621940} with intermediate result usage. 
To employ our module, the user can express the algorithm using \texttt{Flink} dataflow programming primitives.


\section{Evaluation}\label{sec:evaluation}

The source of \name is available online.\footnote{\url{https://fenix.tecnico.ulisboa.pt/homepage/ist162460/veilgraph}} 
We provide an API allowing programmers to implement their logic succinctly.
\name was evaluated with the PageRank power method algorithm~\cite{pageRank}.
The PageRank logic is succinctly implemented as a function as follows:
\begin{lstlisting}[
    basicstyle=\fontsize{8}{9}\ttfamily, %or \small or \footnotesize etc.
    language={Java},
]
public static class PageRankFunction implements Function<MessageIterator<Double>, Double>, Serializable {

        private final Double dampening;

        public PageRankFunction(Double dampening) {
            this.dampening = dampening;
        }
        
        @Override
        public Double apply(final MessageIterator<Double> inMessages) {

            double rankSum = 0.0;
            for (double msg : inMessages) {
                rankSum += msg;
            }
            return (this.dampening * rankSum) + (1 - this.dampening);
        }
    }
\end{lstlisting}
This is then passed on to the underlying graph processing paradigm, as such:

\begin{lstlisting}[
    basicstyle=\fontsize{8}{9}\ttfamily, %or \small or \footnotesize etc.
    language={Java},
]
PageRankFunction prf = new PageRankFunction
(dampeningFactor);

GraphAlgorithm<Long, Double, Double, DataSet<Tuple2<Long, Double>>> algo = new VertexCentricAlgorithm
(iterations, prf);

DataSet<Tuple2<Long, Double>> ranks = summaryGraph.run(algo);
\end{lstlisting}

While we focus our evaluation on PageRank, we note that other random walk based algorithms can be expressed easily.

In our PageRank implementation, all vertices are initialized with the same value at the beginning.
We focus on a vertex-centric implementation of PageRank, where for each iteration, each vertex $u$ sends its value (divided by its outgoing degree) through each of its outgoing edges.
A vertex $v$ defines its score as the sum of values received from its incoming edges, multiplied by a constant factor $\beta$ and then summed with a constant value $(1 - \beta)$ with $0 \leq \beta \leq 1$.
PageRank, based on the random surfer model, uses $\beta$ as a dampening factor.
For our work, this means that whether one considers one-time offline processing or online processing over a stream of graph updates, the underlying computation of PageRank is an approximate numerical version well known in the literature.

This distinction is important, for when we state \name enables approximate computing, we are considering a potential for applicability to a scope of graph algorithms, such as algorithms for computing eigenvector centrality and optimization algorithms for finding communities in networks.
Whether the specific graph algorithm itself incurs numerical approximations (such as the \textit{power method}) or not, that is orthogonal to our model.

\textbf{Experiments.}
Our experiments included two setups.
Most experiments were performed on an SMP machine with 256 GB RAM and 8 Intel(R) Xeon(R) CPU E7- 4830 @ 2.13GHz with eight cores each.
Each dataset execution was performed with a parallelism of four\footnote{\url{https://ci.apache.org/projects/flink/flink-docs-stable/dev/parallel.html}}, meaning a single \texttt{JobManager} and four \texttt{TaskManagers} (master and worker nodes in \texttt{Flink}).
The workers were set to use either 4GB or 8GB of memory, depending on the dataset.

To realistically evaluate the effect that cluster execution has on speedup, we also evaluated one of the larger  datasets of our experiments (\texttt{amazon-2008}) in Google Cloud clusters of progressive size.
This aspect of our evaluation is further detailed in our description of the \texttt{amazon-2008} experiments.

\textit{Application Scenario.} In our scenario, PageRank is initially computed over the complete graph $G$ and then processing a stream $S$ of chunks of incoming edge updates in \name.
For each chunk received in \name we: \textit{1)} integrate the edge updates into the graph; \textit{2)} compute the summarized graph $\mathcal{G} = (\mathcal{V}, \mathcal{E})$ as described in Section~\ref{subsec:building_the_model} and execute PageRank over $\mathcal{G}$.
Thus, we \textit{process} a query when PageRank  (summarized or complete) is executed after integrating a sizeable chunk of updates, i.e. not after each individual update, thus favouring (non-approximate) \texttt{Flink} for comparability (otherwise, although exactly accurate, it is much slower).

To reduce variability, the stream $S$ of edge updates was set up so the number $Q$ of queries for each dataset and parameter combination is always fixed: fifty ($Q = 50$).
For each dataset and stream size, we defined (offline) a file containing the stream of edge updates.
Additionally, for each dataset, streams were generated by uniformly sampling from the edges in the original dataset file.
A stream size of $|S| = 40000$ was used, equaling a chunk of 800 edges added before executing every query.

\textbf{On stream size.}
The number of stream queries $Q$ and number of edge additions per query update were chosen to favour (non-approximate) \texttt{Flink} for comparability in a sensible scenario against summarized executions.
For $|S| = 40000$, if we added 8 edges instead of 800 before executing each query, we would have $Q = 5000$.
This is a much longer sequence of queries, where the graph barely changes between them, with \name  having near-zero execution times in most, where \texttt{Flink} would be 100-fold slower processing the complete graph.
To avoid that, we chose empirically the value of 800 edges before each query (resulting $Q$=50).

We test both edge additions and deletions.
Every time we add edges, we remove an amount equal to 20\% of the number of edges added.
The edges to remove are chosen at random with equal probability.
When additions and removals are applied before an execution, the removal only targets remaining edges which already existed in the original graph or that were added in an older update that preceded a previous execution.

For each dataset and stream $S$ of size $Q$, each combination of parameters $r, n, \Delta$ is tested against a replay of the same stream.
Essentially, each execution (representing a unique combination of parameters) will begin with a complete PageRank execution followed by $Q = 50$ summarized PageRank executions.
This initial computation represents the real-world situation where the results have already been calculated for the whole graph.
In such a situation, one is focused on the incoming updates.
For each dataset and stream $S$, we also execute a scenario which does not use the parameters: it starts likewise with a complete execution of PageRank, but the complete PageRank is executed for all $Q$ queries.
This is required to obtain ground-truth results to measure accuracy and performance of the summarized implementation of the model.
Many datasets such as web graphs are usually provided in an incidence model~\cite{BoVWFI}.
In this model, the out-edges of a vertex are provided together sequentially.
This may lead to an unrealistically favorable scenario, as it is a property that will not necessarily hold in online graphs and which may benefit performance measurements.
To account for this fact, we tested the same parameter combinations with a previous shuffling of stream $S$ (a single shuffle was performed offline a priori so that the randomized stream is the same for different parameter $r, n, \Delta$ combinations that were tested).
This increases the entropy and allows us to validate our model under fewer  assumptions.

\subsection{Datasets}\label{sec:evaluation:datasets}

The datasets' vertex and edge counts are shown in Table~\ref{table:datasets}.
We evaluate results over two types of graphs: web graphs and social networks.
The web graphs and social networks were obtained from the Laboratory for Web Algorithmics~\cite{BRSLLP},~\cite{BoVWFI}.
These datasets were used to evaluate the model against different types of real-world networks.

\begin{table}
	\centering
	\caption{Datasets from the Laboratory for Web Algorithmics~\cite{BoVWFI}. Web graphs are indicated with $^1$ and social networks with $^2$.}
	\begin{tabular}{|>{\centering\arraybackslash} m{2.5cm}|>{\centering\arraybackslash} m{1.5cm}|>{\centering\arraybackslash} m{1.9cm}|>{\centering\arraybackslash} m{1.0cm}|}\hline
	
	Dataset & $|V|$ & $|E|$\\ \hline
	\texttt{eu-2005}$^1$ & 862,664 & 19,235,140\\ \hline
	\texttt{dblp-2010}$^2$ & 326,186 & 1,615,400\\ \hline  
	\texttt{amazon-2008}$^2$ & 735,323 & 5,158,388\\ \hline

	\end{tabular}
	\label{table:datasets}
\end{table}

\subsection{Assessment Metrics}\label{subsec:assessment-metrics}

We measure the results of our approach in terms of: \textit{a)} ability to delay computation in light of result accuracy (top graph of each figure); \textit{b)} obtained execution speedup  (middle graph of each figure); \textit{c)} reduction in number of processed edges (bottom graph of each figure).
Accuracy in our case takes on special importance and requires additional attention to detail.
The PageRank score itself is a measure of importance and we wish to compare rankings obtained on a summarized execution against rankings obtained on the non-summarized graph.
As such, what is desired is a method to compare rankings.

Rank comparison can incur different pitfalls.
If we order the list of PageRank results in decreasing order, only a set of top-vertices is relevant.
After a given index in the ranking, the centrality of the vertices is so low that they are not worth considering for comparative purposes.
But where to define the truncation?
The decision to truncate at a specific position of the rank is arbitrary and leads to the list being incomplete.
Furthermore, the contention between ranking positions is not constant.
Competition is much more intense between the first and second-ranked vertices than between the two-hundredth and two-hundredth and first.

We employed Rank-Biased-Overlap (RBO)~\cite{Webber:2010:SMI:1852102.1852106} as a meaningful evaluation metric (representing relative accuracy) developed to deal with these inherent issues of rank comparison.
RBO has useful properties such as weighting higher ranks more heavily than lower ranks, which is a natural match for PageRank as a vertex centrality measure.
It can also handle rankings of different lengths.
This is in tune with the output of a centrality algorithm such as PageRank.
The RBO value obtained from two rank lists is a scalar in the interval $[0, 1]$.
It is zero if the lists are completely disjoint and one if they are completely equal.
While more recent comparison metrics have been proposed~\cite{moffat2018computing}, they go beyond the scope of what is required in our comparisons.
The quality of our case study algorithm's accuracy is itself produced by a comparison (between sequences of rank lists).
As far as we know, and due to the specificity of our evaluation, there is no better-suited baseline in the literature against which to compare our ranking comparison results.

Performance-wise, we test values of $r$ associated to different levels of sensitivity to vertex degree change (the higher the number, the less expected objects to process per query).
With $n = 0$, we minimize the expansion around the first set so that $K_{n} = \varnothing$ and $K_{r} \cup K_{n} = K_{r}$.
For $n = 1$, we are taking a more conservative approach regarding result accuracy.
An overall tendency to expect is that the higher the value of $n$ is, the higher the RBO (this is demonstrated in our results).
The $\Delta$ values were chosen to evaluate individual different weight schemes applied to vertex score changes.
The relation between parameters $r$ and $n$ has a greater impact in performance and accuracy than the relation of any of these parameters with $\Delta$.
We tested with two sets of parameter combinations: 
\begin{itemize}
    \item RBO-oriented ($r = 0.05, n = 2, \Delta = 1.0$), ($r = 0.05, n = 2, \Delta = 0.5$), ($r = 0.05, n = 6, \Delta = 1.0$), ($r = 0.05, n = 6, \Delta = 0.5$).
    This has a very low threshold of sensitivity to the ratio of vertex degree change ($r = 0.05$).
    
    \item Performance-oriented ($r = 0.20, n = 0, \Delta = 0.5$), ($r = 0.20, n = 0, \Delta = 1.0$), ($r = 0.20, n = 1, \Delta = 0.5$), ($r = 0.20, n = 1, \Delta = 1.0$), ($r = 0.20, n = 4, \Delta = 1.0$).
    With $r = 0.20$, the goal is to be less sensitive pertaining degree change ratio.
\end{itemize}
For both of these combinations, we test with low and high values of $n$ to examine how expanding the neighborhood of vertices complements the initial degree change ratio filter.

Using a higher number of ranks for the RBO evaluation favors a comparison of calculated ranks which has greater resolution, as more vertices are being compared.
In our evaluation, the RBO of each execution is calculated using 10\%  of the complete graph's vertex count as the number of top ranks to compare.\footnote{
Every 10 executions, we calculate RBO using all of the vertices of the graph  periodically ensuring no artifacts are masked in the lower rank values.}

\subsection{Results}\label{subsec:results}



Results are tailored to illustrate the impact of the \name model based on three metrics: \textit{a)} obtained speedup for each execution (compared against the execution over the complete graph); \textit{b)} RBO evolution as the number of executions increases (it starts with a value of 1 as the complete version of PageRank is executed initially for all execution strategy scenarios); \textit{c)} summary graph edge count as a percentage of the original graph's edge count.

For these three metric categories, we present the best-three and worst-three results obtained within each category.
This means that the parameters ($r, n, \Delta$) producing the best accuracy result are not necessarily the same ones producing the best speedups.
\textbf{The horizontal axis represents the same for all plots}: it is the sequence of queries from 1 up to $Q = 50$.
Each figure has three graphics in a column, corresponding to speedup (top), accuracy measured with RBO (middle) and summary graph edge count as percentage of the complete graph edge count (bottom).
Due to how the dynamics of parameter combinations and the structure of the data sets behave, some parameter combinations produced extremely similar values, leading to almost overlapping plots.
We describe first the meaning of the results observed for the web graph and \texttt{eu-2005} and last for the social graphs \texttt{dblp-2010} and \texttt{amazon-2008}.

One needs to take into account this is a challenging assessment context for \titlename. In fact, between each consecutive pair of the 50 queries (i.e., on every of the 800 edge/vertex updates we are ingesting between them), if the user prompted a query execution, \titlename could offer near-instant results against the previously summarized graph, contrary to a full graph execution (thus yielding several 100-fold speedups each time), and still provide results with very high RBO (in line with those from the preceding and successor of the pair of queries where the update lied between).

\setlength{\abovecaptionskip}{2pt} 
\newcommand{\figlen}{0.475} 

\begin{figure}
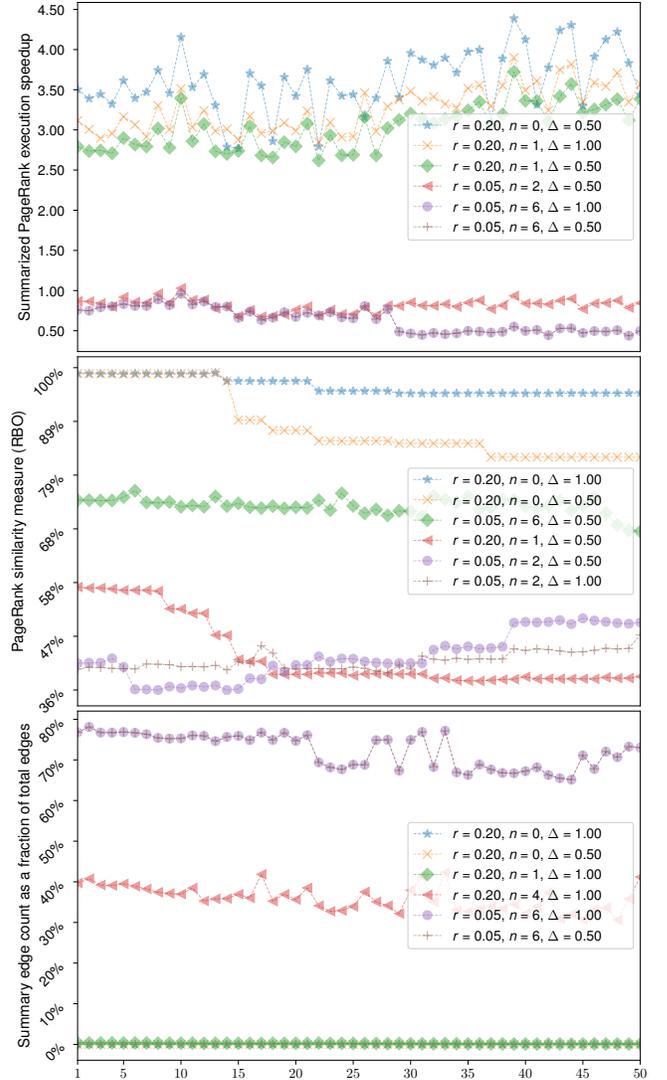

    \subfloat{\scalebox{\figlen}{\input{eu-2005-40000-random_30_86266_P4_0.85_D-Time-vs-exact-speedup-top-3-bottom-3_nolabels.pgf}}}
    \newline
    \vskip-1.74cm
    \subfloat{\scalebox{\figlen}{\input{eu-2005-40000-random_30_86266_P4_0.85_D-RBO-top-3-bottom-3_nolabels.pgf}}}
    \newline
    \vskip-1.74cm
    \subfloat{\scalebox{\figlen}{\input{eu-2005-40000-random_30_86266_P4_0.85_D-Savings-edge-top-3-bottom-3.pgf}}}
    \caption{\texttt{eu-2005}. Top-3/Bottom-3 speedups (upper). Top-3/Bottom-3 RBO results (middle). Top-3/Bottom-3 summary graph edge savings (bottom). Horizontal axis is the sequence of $Q = 50$ queries.}
    \label{fig:eu-2005-40000-random}
\end{figure}

\textbf{\texttt{eu-2005}}: 
Results for this dataset are on Figure~\ref{fig:eu-2005-40000-random}.
Speedups of around 3.00 were achieved (see the blue star and yellow cross markers at the top of Figure~\ref{fig:eu-2005-40000-random}).
These are parameter combinations which promote speed, only considering for the \textit{hot vertex} set the vertices whose degree changed by at least 20\%.

Combinations with $r = 0.20$ and $n = 0$ achieved the best RBO throughout all executions for this dataset.
This was due to the same phenomenon which arose with dataset \texttt{cnr-2000}.

The parameters with more conservative values (bigger $n$ and lower $r$) led to the biggest summary graphs.
However, these were not the ones producing the highest RBO values throughout executions.

\begin{figure}
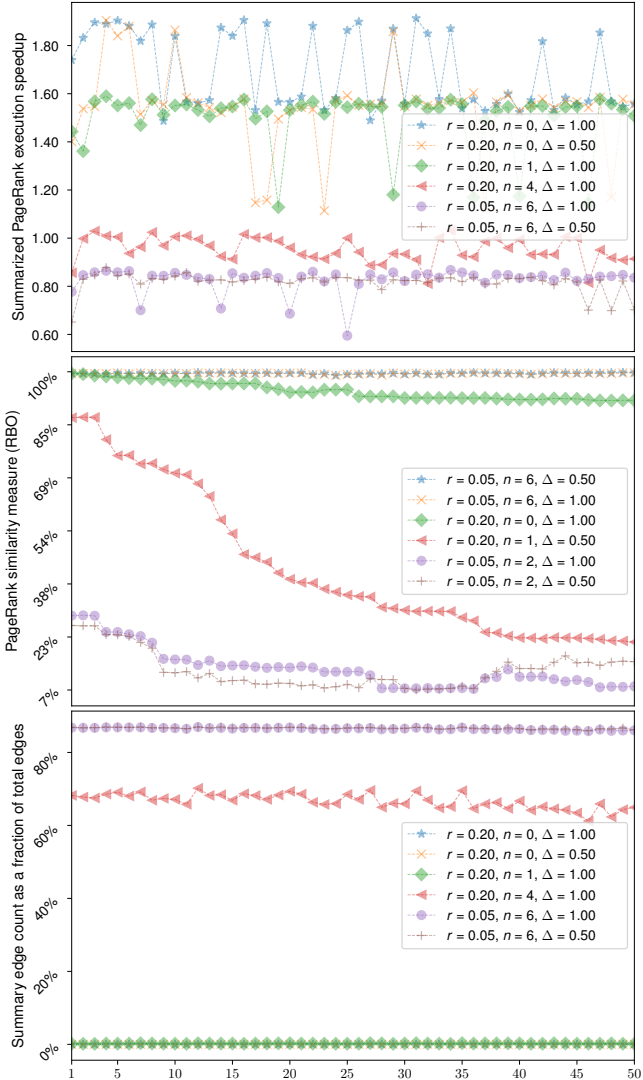

    \subfloat{\scalebox{\figlen}{\input{dblp-2010-40000-random_30_32186_P4_0.85_D-Time-vs-exact-speedup-top-3-bottom-3_nolabels.pgf}}}
    \newline
    \vskip-1.74cm
    \subfloat{\scalebox{\figlen}{\input{dblp-2010-40000-random_30_32186_P4_0.85_D-RBO-top-3-bottom-3_nolabels.pgf}}}
    \newline
    \vskip-1.74cm
    \subfloat{\scalebox{\figlen}{\input{dblp-2010-40000-random_30_32186_P4_0.85_D-Savings-edge-top-3-bottom-3.pgf}}}
    \caption{\texttt{dblp-2010}. Top-3/Bottom-3 speedups (upper). Top-3/Bottom-3 RBO results (middle). Top-3/Bottom-3 summary graph edge savings (bottom). Horizontal axis is the sequence of $Q = 50$ queries.}
    \label{fig:dblp-2010-40000-random}
\end{figure}

\textbf{\texttt{dblp-2010}}:
Results are shown in Figure~\ref{fig:dblp-2010-40000-random}.
The best speedups obtained were around 1.60-1.80 for high values of the update ratio threshold ($r = 0.20$) and lower levels of neighborhood expansion ($n = 0$, $n = 1$).
These are the markers with blue stars, yellow crosses and green diamonds.
For result accuracy, parameter combination ($r = 0.20, n = 1, \Delta = 0.50$) with the red left-facing triangles started at around 85\% and decreased steadily as executions progressed.
Adjusting this combination by switching $n$ from $0$ to $1$ produced a plot with RBO values above 90\%, shown in the green diamond marker plot.
The best RBO values were produced by the bigger value of $n = 6$.
These same parameter combinations using $r = 0.05$ and $n = 6$ also led to a summary graph edge count very close to the complete graph's edge count.
A more balanced combination (slightly lower $n$ and higher $r$) produced a summary graph whose edge fraction (with respect to the total graph) hovered around 70\%.

\begin{figure}
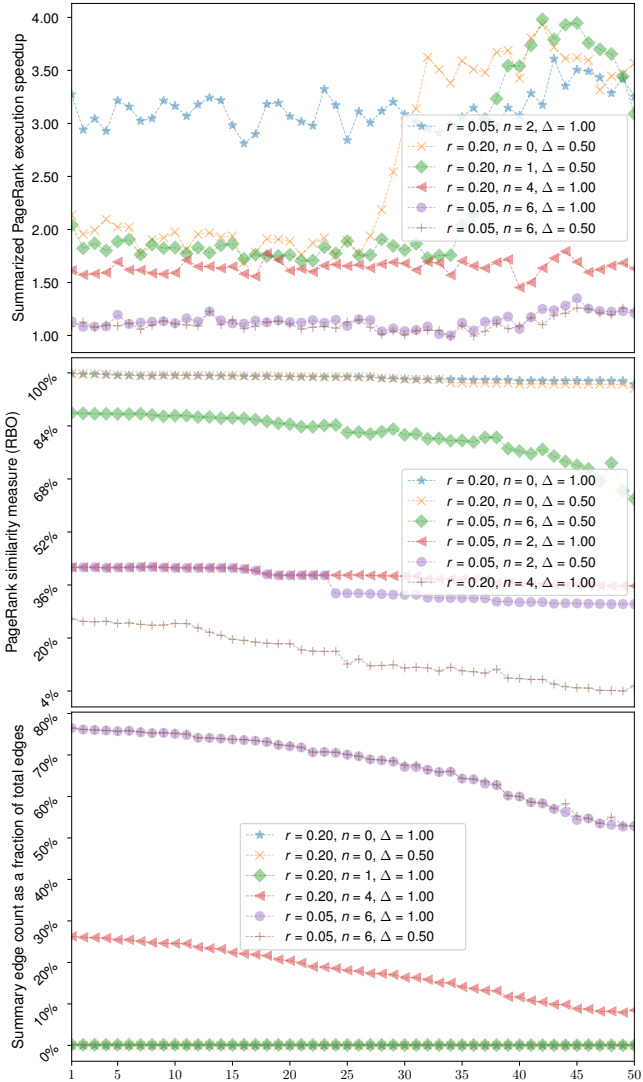

    \subfloat{\scalebox{\figlen}{\input{amazon-2008-40000-random_30_73532_P4_0.85_D-Time-vs-exact-speedup-top-3-bottom-3_nolabels.pgf}}}
    \newline
    \vskip-1.74cm
    \subfloat{\scalebox{\figlen}{\input{amazon-2008-40000-random_30_73532_P4_0.85_D-RBO-top-3-bottom-3_nolabels.pgf}}}
    \newline
    \vskip-1.74cm
    \subfloat{\scalebox{\figlen}{\input{amazon-2008-40000-random_30_73532_P4_0.85_D-Savings-edge-top-3-bottom-3.pgf}}}
    \caption{\texttt{amazon-2008}. Top-3/Bottom-3 speedups (upper). Top-3/Bottom-3 RBO results (middle). Top-3/Bottom-3 summary graph edge savings (bottom). Horizontal axis is the sequence of $Q = 50$ queries.}
    \label{fig:amazon-2008-40000-random}
\end{figure}

\textbf{\texttt{amazon-2008}}: 
At the top of Figure~\ref{fig:amazon-2008-40000-random} we have speedups of around 3.00.
Parameter combination ($r = 0.05, n = 2, \Delta = 1.00$) with the blue star plot had a stable speedup value close to this, only to be surpassed (starting from execution $30$) by parameter combinations ($r = 0.20, n = 0, \Delta = 0.50$) and ($r = 0.20, n = 1, \Delta = 0.50$), represented by the yellow cross and the green diamond markers respectively.

This dataset exhibited an unexpected tendency where higher RBO values were retained for a mix of $r = 0.20$ and $n = 0$ with varying values of $\Delta$.
These accuracy results are contrasted by the lower value of RBO for ($r = 0.05, n = 6, \Delta = 0.50$).
We attribute the observed behavior of lower RBO with a much higher $n$ to the impact of the edge removal on the topology of the \texttt{amazon-2008} dataset.

The number of summary graph edges as a fraction of the complete graph's edges is in line with previous results: greater values of $n$ led to a bigger summary graph.
Still, there was a pattern with $n = 6$ where there was a tendency for RBO to decrease (green diamond marker on the middle graph) coupled with a tendency for the summary graph edge fraction to decrease (purple circle marker on the bottom graph) too.

\begin{figure}
	\centering
	\includegraphics[width=0.9\linewidth]{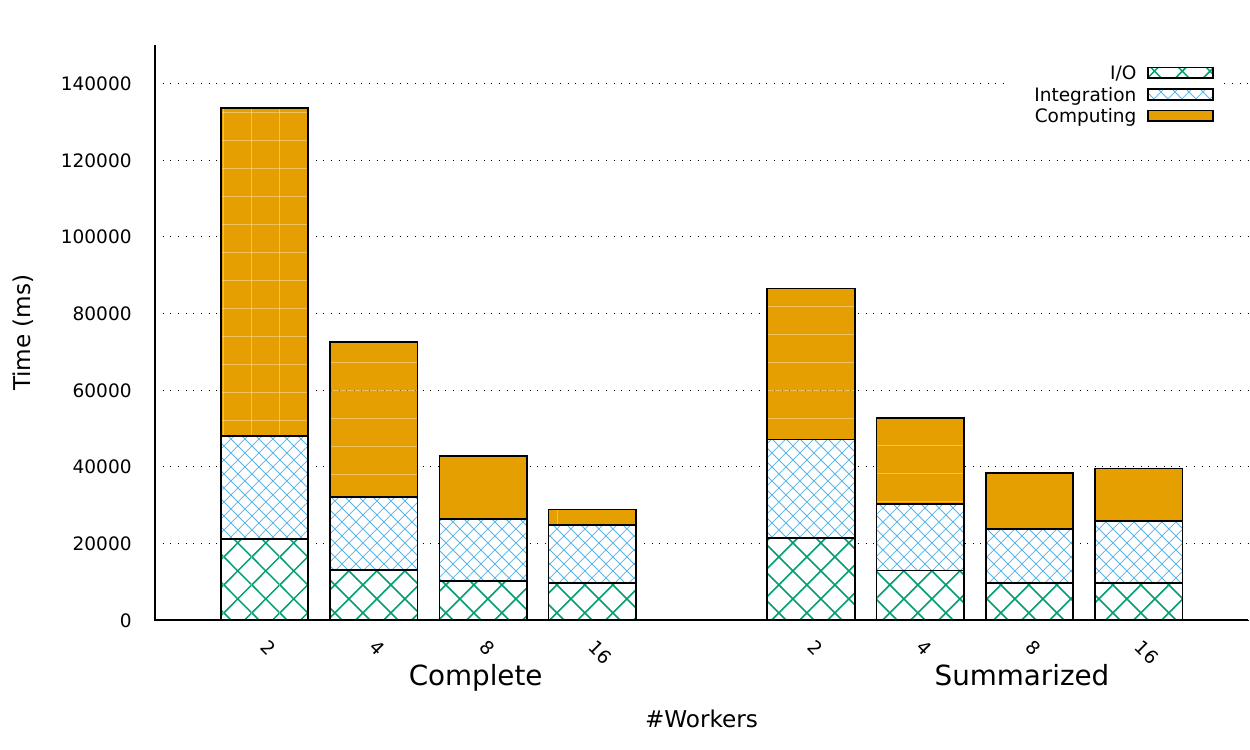}
	\caption{Separate measurement types during execution of \texttt{amazon-2008}.}
	\label{fig:amazon-separate-times}
\end{figure}

For this dataset we also present in Fig.~\ref{fig:amazon-separate-times} a breakdown of the way execution time is distributed across different phases of our model.
Effectively, \texttt{Flink's} job chaining for dataset reuse requires intermediate checkpoints of write/read I/O activity (whether this is in a local disk or in distributed storage) for both integrating updates (for either complete or summarized executions) into the graph as well as building the summary big vertex.
The measurements in Fig.~\ref{fig:amazon-separate-times} were obtained for Google Cloud clusters with a number of workers of 2, 4, 8 and 16.
In the image we see the breakdown of three categories of time consumption:

\begin{itemize}
\item Column bottom: time spent with writing and ingesting the graph back from distributed storage.
\item Column middle: time consumed with \texttt{Flink}-specific internal \texttt{DataSet} construction and initial distribution.
\item Column top: actual computing time.
\end{itemize}

\textbf{Discussion.}
The speedups indicative that as we test with larger datasets, the benefit of executing a graph algorithm over just \name's $K$ set instead of the complete graph is beneficial to performance while maintaining very competitive levels of accuracy, such as the case of parameters ($r = 0.20, n = 0, \Delta = 0.50$), achieving over 50\% faster computational time with an RBO above 90\%.

Despite this, our method has achieved a very good trade-off between result accuracy and reduction in total computation (be it in number of processed graph elements or direct time comparison), bearing in mind the lower-bound performance context we are assessing (recall: we are executing and comparing with the full processing of the graph just on 50 instants, against doing it on every vertex/edge update).

Randomized edge removals were used to assess the robustness of the model.
Removals have an impact on graph topology which does not necessarily manifest as one would expect.
Extending the computational scope by using conservative parameters in \name overall promoted a greater accuracy in our experiments, but special consideration must be given to the cases where removals may lead to a cascading effect, where extending the scope of the model promotes a more intense error propagation, which is undesirable.

The visualization of separate computational times as shown in Fig.~\ref{fig:amazon-separate-times} is revealing of a bi-dimensional cost that is often overlooked in these distributed computational platforms.
This cost manifests, as one would expect, in the overheads of communication between cluster elements and the writing/reading of data from distributed storage.
Its second dimension (also a trait in \texttt{Apache Spark}) is located in the runtime of \texttt{Flink}.
The user-logic (in our case written in \texttt{Java}) for the most part is able to abstract away that the underlying execution will take place in a distributed system.
However, un-optimized, this incurs relevant communication costs which are not always obvious and whose measurement requires breaking down this abstraction.
This may be achieved by directly measuring the metrics of operators executed as part of the optimized execution plan produced by the \texttt{Flink} compiler and is something we plan to explore in future work.
As proof, the network and integration time for the summarized version actually increased from 8 to 16 workers (column middle).
Another interesting observation is that with more than 8 workers, the speedup of the summarized version was lower than in the complete version.
This is due to the summary version processing fewer  elements and thus not leveraging  extra workers so much.


\section{Related Work}\label{sec:related}
Our multidisciplinary work encompasses: paradigms to express graph computations; stream processing; approximation techniques.
We present the most relevant state-of-the-art contributions in this respect.

\textbf{Kineograph.}
A distributed system to capture the relations in incoming data feeds~\cite{Cheng:2012:KTP:2168836.2168846}, built to maintain timely updates against a continuous flux of data.
Its architecture uses \textit{ingest} nodes to register graph update operations as identifiable transactions, which are then distributed to \textit{graph} nodes.
Nodes of the later type form a distributed in-memory key/value store.
KineoGraph performs computation on static snapshots, which simplifies algorithm design. 
\name goes beyond this by giving users the flexibility to design algorithms for either the complete graph or summarized versions, with little difference.
Users can incorporate the awareness that the graph has changed, or opt to design an algorithm that considers the current graph as a static version, like KineoGraph.

\textbf{KickStarter.}
A technique for trimming approximate values of vertex subsets which were impacted by edge deletion~\cite{Vora:2017:KFA:3037697.3037748}.ithm  logic  and  als
KickStarter deals with edge deletions this by identifying values impacted by the deletions and adapting the network impacts before the following computation, achieving good results on real-world use-cases.
By focusing on monotonic graph algorithms, its scope is narrowed to selection-based algorithms.
We decouple in \name the approximation technique, the summarization model and the algorithm type.
Thus, we are able to offer the big vertex model and provide a structured sequence of steps to integrate another model or approximation technique (e.g. KickStarter's own technique could be a candidate).

\textbf{Tornado.}
A system with an asynchronous bounded iteration model, offering fine-grained updates while ensuring correctness~\cite{Shi:2016:TSR:2882903.2882950}. 
It is based on the observations that: \textit{1)} loops starting from \textit{good enough} guesses usually converge quickly; \textit{2)} for many iterative methods, the running time is closely relative to the approximation error.
Whenever a result request is received, the model creates a branch loop from the main loop.
This branch loop uses the most recent approximations as a guess for the algorithm.
It is a technique that could benefit from applying \name's summarization model.
Tornado's main loop could produce approximations faster, making  \textit{good guesses} readily available as  queries arrive.

\textbf{Naiad.}
A dataflow processing system~\cite{murray2013naiad} offering different levels of complexity and abstractions to programmers.
It allows programmers to implement graph algorithms such as Weakly Connected Components (WCC), Approximate Shortest Paths (ASP) and others while achieving better performance than other systems.
Naiad allows programmers to use common high-level APIs to express algorithm logic and also offers a low-level API for performance.
Its concepts are important and other systems that could benefit from offering tiered programming abstraction levels.


\vspace{-6pt}
\section{Conclusion}\label{sec:conclusion}

We designed \name and evaluated its approximate computing capabilities over PageRank, a type of algorithm based on random walk making use of the graph summary representation that we propose.
It provides a well-defined structure to incorporate custom approximate processing strategies and to enable choosing between built-in behaviors.
Our experiments in the context of random walk problems lead us to conclude that the \name model, even when tested in a challenging context for comparability, is a viable basis to enable faster, more efficient and configurable graph processing on this type of problem.
The results we obtain were produced under a stream scenario where graph updates are big enough so that the changes to the graph do not explicitly benefit the summarization model we evaluated.
We plan to further research the challenge of the disruptive  aspects of edge deletions over graph topology and how that may speed up or slow down the propagation of approximation errors.

\vspace{-6pt}
\bibliographystyle{IEEEtran} 
\bibliography{references}

\begin{thebibliography}{10}
\providecommand{\url}[1]{#1}
\csname url@samestyle\endcsname
\providecommand{\newblock}{\relax}
\providecommand{\bibinfo}[2]{#2}
\providecommand{\BIBentrySTDinterwordspacing}{\spaceskip=0pt\relax}
\providecommand{\BIBentryALTinterwordstretchfactor}{4}
\providecommand{\BIBentryALTinterwordspacing}{\spaceskip=\fontdimen2\font plus
\BIBentryALTinterwordstretchfactor\fontdimen3\font minus
  \fontdimen4\font\relax}
\providecommand{\BIBforeignlanguage}[2]{{%
\expandafter\ifx\csname l@#1\endcsname\relax
\typeout{** WARNING: IEEEtran.bst: No hyphenation pattern has been}%
\typeout{** loaded for the language `#1'. Using the pattern for}%
\typeout{** the default language instead.}%
\else
\language=\csname l@#1\endcsname
\fi
#2}}
\providecommand{\BIBdecl}{\relax}
\BIBdecl

\bibitem{pageRank}
\BIBentryALTinterwordspacing
L.~Page, S.~Brin, R.~Motwani, and T.~Winograd, ``{The PageRank Citation
  Ranking: Bringing Order to the Web.}'' Stanford InfoLab, Technical Report
  1999-66, 1999. [Online]. Available:
  \url{http://ilpubs.stanford.edu:8090/422/}
\BIBentrySTDinterwordspacing

\bibitem{GraphSampling2013}
\BIBentryALTinterwordspacing
P.~Hu and W.~C. Lau, ``{A Survey and Taxonomy of Graph Sampling},''
  \emph{CoRR}, vol. abs/1308.5, aug 2013. [Online]. Available:
  \url{https://arxiv.org/abs/1308.5865}
\BIBentrySTDinterwordspacing

\bibitem{ProbabilisticAccuracy2006}
\BIBentryALTinterwordspacing
M.~Rinard, ``{Probabilistic Accuracy Bounds for Fault-tolerant Computations
  That Discard Tasks},'' in \emph{Proceedings of the 20th Annual International
  Conference on Supercomputing}, ser. ICS '06.\hskip 1em plus 0.5em minus
  0.4em\relax New York, NY, USA: ACM, 2006, pp. 324--334. [Online]. Available:
  \url{http://doi.acm.org/10.1145/1183401.1183447}
\BIBentrySTDinterwordspacing

\bibitem{LoadShedding2003}
\BIBentryALTinterwordspacing
B.~B. Babcock, M.~Datar, R.~Motwani, B.~B. Mayur, B.~B. Babcock, M.~Datar, and
  R.~Motwani, ``{Load Shedding Techniques for Data Stream Systems},'' in
  \emph{In Proc. of the 2003 Workshop on Management and Processing of Data
  Streams (MPDS}, 2003, pp. 1--3. [Online]. Available:
  \url{http://citeseerx.ist.psu.edu/viewdoc/summary?doi=10.1.1.5.1941}
\BIBentrySTDinterwordspacing

\bibitem{LinkEvolutions}
\BIBentryALTinterwordspacing
S.~Chien, C.~Dwork, R.~Kumar, D.~R. Simon, and D.~Sivakumar, ``{Link
  Evolutions: Analysis and Algorithms},'' \emph{Internet Math.}, vol.~1, no.~3,
  pp. 277--304, 2003. [Online]. Available:
  \url{http://projecteuclid.org/euclid.im/1109190963}
\BIBentrySTDinterwordspacing

\bibitem{Knowing2014}
\BIBentryALTinterwordspacing
S.~Agarwal, H.~Milner, A.~Kleiner, A.~Talwalkar, M.~Jordan, S.~Madden,
  B.~Mozafari, and I.~Stoica, ``{Knowing when You'Re Wrong: Building Fast and
  Reliable Approximate Query Processing Systems},'' in \emph{Proceedings of the
  2014 ACM SIGMOD International Conference on Management of Data}, ser. SIGMOD
  '14.\hskip 1em plus 0.5em minus 0.4em\relax New York, NY, USA: ACM, 2014, pp.
  481--492. [Online]. Available:
  \url{http://doi.acm.org/10.1145/2588555.2593667}
\BIBentrySTDinterwordspacing

\bibitem{kalavri2016shortest}
V.~Kalavri, T.~Simas, and D.~Logothetis, ``The shortest path is not always a
  straight line: leveraging semi-metricity in graph analysis,''
  \emph{Proceedings of the VLDB Endowment}, vol.~9, no.~9, pp. 672--683, 2016.

\bibitem{Langville:2004:UPI:1013367.1013491}
\BIBentryALTinterwordspacing
A.~N. Langville and C.~D. Meyer, ``Updating pagerank with iterative
  aggregation,'' in \emph{Proceedings of the 13th International World Wide Web
  Conference on Alternate Track Papers \&Amp; Posters}, ser. WWW Alt.
  '04.\hskip 1em plus 0.5em minus 0.4em\relax New York, NY, USA: ACM, 2004, pp.
  392--393. [Online]. Available:
  \url{http://doi.acm.org/10.1145/1013367.1013491}
\BIBentrySTDinterwordspacing

\bibitem{Cheng:2012:KTP:2168836.2168846}
\BIBentryALTinterwordspacing
R.~Cheng, J.~Hong, A.~Kyrola, Y.~Miao, X.~Weng, M.~Wu, F.~Yang, L.~Zhou,
  F.~Zhao, and E.~Chen, ``Kineograph: Taking the pulse of a fast-changing and
  connected world,'' in \emph{Proceedings of the 7th ACM European Conference on
  Computer Systems}, ser. EuroSys '12.\hskip 1em plus 0.5em minus 0.4em\relax
  New York, NY, USA: ACM, 2012, pp. 85--98. [Online]. Available:
  \url{http://doi.acm.org/10.1145/2168836.2168846}
\BIBentrySTDinterwordspacing

\bibitem{Junghanns:2016:AEP:2980523.2980527}
\BIBentryALTinterwordspacing
M.~Junghanns, A.~Petermann, N.~Teichmann, K.~G\'{o}mez, and E.~Rahm,
  ``Analyzing extended property graphs with apache flink,'' in
  \emph{Proceedings of the 1st ACM SIGMOD Workshop on Network Data Analytics},
  ser. NDA '16.\hskip 1em plus 0.5em minus 0.4em\relax New York, NY, USA: ACM,
  2016, pp. 3:1--3:8. [Online]. Available:
  \url{http://doi.acm.org/10.1145/2980523.2980527}
\BIBentrySTDinterwordspacing

\bibitem{Malewicz:2010:PSL:1807167.1807184}
\BIBentryALTinterwordspacing
G.~Malewicz, M.~H. Austern, A.~J. Bik, J.~C. Dehnert, I.~Horn, N.~Leiser, and
  G.~Czajkowski, ``Pregel: A system for large-scale graph processing,'' in
  \emph{Proceedings of the 2010 ACM SIGMOD International Conference on
  Management of Data}, ser. SIGMOD '10.\hskip 1em plus 0.5em minus 0.4em\relax
  New York, NY, USA: ACM, 2010, pp. 135--146. [Online]. Available:
  \url{http://doi.acm.org/10.1145/1807167.1807184}
\BIBentrySTDinterwordspacing

\bibitem{Kalavri:2014:ALG:2621934.2621940}
\BIBentryALTinterwordspacing
V.~Kalavri, S.~Ewen, K.~Tzoumas, V.~Vlassov, V.~Markl, and S.~Haridi,
  ``Asymmetry in large-scale graph analysis, explained,'' in \emph{Proceedings
  of Workshop on GRAph Data Management Experiences and Systems}, ser.
  GRADES'14.\hskip 1em plus 0.5em minus 0.4em\relax New York, NY, USA: ACM,
  2014, pp. 4:1--4:7. [Online]. Available:
  \url{http://doi.acm.org/10.1145/2621934.2621940}
\BIBentrySTDinterwordspacing

\bibitem{BoVWFI}
P.~Boldi and S.~Vigna, ``The {W}eb{G}raph framework {I}: {C}ompression
  techniques,'' in \emph{Proc. of the Thirteenth International World Wide Web
  Conference (WWW 2004)}.\hskip 1em plus 0.5em minus 0.4em\relax Manhattan,
  USA: ACM Press, 2004, pp. 595--601.

\bibitem{BRSLLP}
P.~Boldi, M.~Rosa, M.~Santini, and S.~Vigna, ``Layered label propagation: A
  multiresolution coordinate-free ordering for compressing social networks,''
  in \emph{Proceedings of the 20th international conference on World Wide Web},
  S.~Srinivasan, K.~Ramamritham, A.~Kumar, M.~P. Ravindra, E.~Bertino, and
  R.~Kumar, Eds.\hskip 1em plus 0.5em minus 0.4em\relax ACM Press, 2011, pp.
  587--596.

\bibitem{Webber:2010:SMI:1852102.1852106}
\BIBentryALTinterwordspacing
W.~Webber, A.~Moffat, and J.~Zobel, ``A similarity measure for indefinite
  rankings,'' \emph{ACM Trans. Inf. Syst.}, vol.~28, no.~4, pp. 20:1--20:38,
  Nov. 2010. [Online]. Available:
  \url{http://doi.acm.org/10.1145/1852102.1852106}
\BIBentrySTDinterwordspacing

\bibitem{moffat2018computing}
A.~Moffat, ``Computing maximized effectiveness distance for recall-based
  metrics,'' \emph{IEEE Transactions on Knowledge and Data Engineering},
  vol.~30, no.~1, pp. 198--203, 2018.

\bibitem{Vora:2017:KFA:3037697.3037748}
\BIBentryALTinterwordspacing
K.~Vora, R.~Gupta, and G.~Xu, ``Kickstarter: Fast and accurate computations on
  streaming graphs via trimmed approximations,'' in \emph{Proceedings of the
  Twenty-Second International Conference on Architectural Support for
  Programming Languages and Operating Systems}, ser. ASPLOS '17.\hskip 1em plus
  0.5em minus 0.4em\relax New York, NY, USA: ACM, 2017, pp. 237--251. [Online].
  Available: \url{http://doi.acm.org/10.1145/3037697.3037748}
\BIBentrySTDinterwordspacing

\bibitem{Shi:2016:TSR:2882903.2882950}
\BIBentryALTinterwordspacing
X.~Shi, B.~Cui, Y.~Shao, and Y.~Tong, ``Tornado: A system for real-time
  iterative analysis over evolving data,'' in \emph{Proceedings of the 2016
  International Conference on Management of Data}, ser. SIGMOD '16.\hskip 1em
  plus 0.5em minus 0.4em\relax New York, NY, USA: ACM, 2016, pp. 417--430.
  [Online]. Available: \url{http://doi.acm.org/10.1145/2882903.2882950}
\BIBentrySTDinterwordspacing

\bibitem{murray2013naiad}
D.~G. Murray, F.~McSherry, R.~Isaacs, M.~Isard, P.~Barham, and M.~Abadi,
  ``Naiad: a timely dataflow system,'' in \emph{Proceedings of the
  Twenty-Fourth ACM Symposium on Operating Systems Principles}.\hskip 1em plus
  0.5em minus 0.4em\relax ACM, 2013, pp. 439--455.

\end{thebibliography}

\end{document}